\documentclass[epj]{svjour}
\usepackage{graphics}
\begin{document}
\title{Random Wave Functions with boundary and normalization constraints:}
\subtitle{Quantum statistical physics meets quantum chaos}
\author{Juan Diego Urbina\inst{1}\fnmsep\thanks{\email{jdurbinag@unal.edu.co}} \and Klaus Richter\inst{2}}
\institute{Grupo de Caos y Complejidad, Departamento de Fisica, Universidad Nacional de Colombia, Bogota, Colombia\and Institut f\"ur Theoretische Physik, Universit\"at Regensburg, 93040 Regensburg, Germany}
\abstract{We present an improved version of Berry's ansatz able to incorporate exactly the existence of boundaries and the correct normalization of the eigenfunction into an ensemble of random waves. We then reformulate the Random Wave conjecture showing that in its new version it is a statement about the universal nature of eigenfunction fluctuations in systems with chaotic classical dynamics. The emergence of the universal results requires the use of both semiclassical methods and a new expansion for a very old problem in quantum statistical physics.
} %end of abstract
\maketitle
\section{Introduction}
\label{intro}
During the Chladni meeting, we have honored Chladni's curiosity about the complex and beautiful patterns visible even in very simple vibrating systems. His interest was rewarded not only with the esthetic pleasure we are sure he experienced when watching this patterns emerge by adding some sand grains to his vibrating plates. Quite likely he also enjoyed applying a deeper geometrical understanding of the modes of vibration (as we would call them now) and ultimately the transformation of sounds into forms. This curiosity is still with us, partially because our modern techniques and tools keep pushing the connection between geometry and vibrations into new domains of science, and partially because some old questions remain unsolved. 

During the conference we have seen examples of both these connections (ranging from mesoscopic physics to the design of musical instruments, or from complex analysis to the theory of visual perception) and these old questions (such as the ones related with the famous ``can you hear the shape of a drum?'').

One connecting link between many of the contributions presented at the conference is the subject of this article. This connecting concept we refer to is a conjecture and a useful practical tool which says that in certain kinds of vibrating systems, universality of the corresponding geometrical patterns appears. Of course, as it stands now, this statement is very loose, and a way to formulate it precisely is one of the open questions we mentioned above. 

Which systems we are talking about and what does universality mean in this context are questions that, thanks to the seminal work of M. V. Berry \cite{berr1}, can be answered by now. However, much is still to be done, and we will ask (and in a couple of cases answer) other subtle aspects of this conjecture.

\section{Berry's conjecture: the random wave model}
\label{sec:1}
In this section we attempt to give a form of the conjecture narrowing both the class of systems and the meaning of universality. We follow the ideas as presented by Berry in his classic paper of 1977 \cite{berr1}. 
\subsection{The universal two-point correlation function}
\label{sec:2}

Consider a quantum particle moving inside a 2D domain (billiard) $\Omega$ with area $A(\Omega)$ and Dirichlet boundary conditions along its boundary $\partial \Omega$. Berry's main observation was that for a given solution $\psi_{n}(\vec{r})$ of the Schr\"odinger equation
\begin{eqnarray}
\nabla^{2}\psi_{n}(\vec{r})&=&-k_{n}^{2}\psi_{n}(\vec{r}), \\
 \psi_{n}(\vec{r})&|&_{\vec{r} \in \partial \Omega}=0, \nonumber
\end{eqnarray}
the spatial two-point correlation function
\begin{equation}
\label{eq:Rn}
R_{n}(\vec{r}_{1},\vec{r}_{2})=\frac{1}{A(S)}\int_{S}\psi_{n}(\vec{r}_{1}+\vec{q})\psi_{n}(\vec{r}_{2}+\vec{q})d\vec{q}
\end{equation}
has a universal form if the classical dynamics inside the billiard is chaotic and the points $\vec{r}_{1},\vec{r}_{2}$ are far from the boundaries. Here $S$ is a subset of $\Omega$ with area $A(S)$ containing many wavelengths. 

The universal two-point correlation function predicted by Berry is given by the well known Bessel function \cite{berr1}
\begin{equation}
\label{eq:bess}
R_{n}(\vec{r}_{1},\vec{r}_{2})=\frac{1}{A(\Omega)}J_{0}(k_{n}|\vec{r}_{1}-\vec{r}_{2}|),
\end{equation}
an extremely robust result which has been extensively checked both numerically and experimentally \cite{sto1}.

There are many arguments (but not a proof) to understand this universality. A useful and straightforward way is to note that in the semiclassical regime the eigenfunctions of classically chaotic quantum billiards should appear structureless, isotropic and roughly homogeneous due to the lack of structure of the classical phase space. This observation can be made precise by assuming that $\psi_{n}(\vec{r})$ is locally written as a superposition of plane waves with fixed wavenumber, directions uniformly distributed and random phases. The ensemble 
\begin{equation}
\label{eq:RWM}
\psi_{n}^{RWM}(x,y;\vec{\delta})=\frac{1}{{\cal N}}\sum_{j}^{L}{\rm e}^{ik_{n}\left[x\cos\left(\frac{2 \pi j}{L}\right)+y\sin\left(\frac{2 \pi j}{L}\right)+\delta_{j}\right]}
\end{equation}
with $\vec{\delta}=(\delta_{1},\ldots,\delta_{L})$ independent random variables uniformily distributed over $(0,2 \pi]$ and ${\cal N}$ a normalization factor is then called a Random Wave Model (RWM).

It can be easily shown that in the limit $L\to \infty$ the RWM correlation function (after average over the random phases) is given again by Eq. (\ref{eq:bess}). This is the fundamental observation made by Berry \cite{berr1}, with very deep and powerful implications: the two-point spatial correlation function of eigenfunctions associated with classically chaotic dynamics (also called ``irregular'') is universal and given by the correlation function of an ensemble of random waves with fixed wavenumber. Note, however, that the two kinds of ensembles are quite different. On one hand, we have the spatial average in Eq. (\ref{eq:Rn}), on the other hand the average over random realizations of the phases in Eq. (\ref{eq:RWM}). In fact, for a given billiard shape, there is a third kind of average, the spectral average, where the fluctuating variables are the values of eigenfunctions with different energy index $n$. This last kind will be of importance later on.

The importance of the relationship between irregular eigenfunctions and random waves goes far beyond this analogy concerning the two-point correlation function. The point is that the semiclassical analysis used by Berry to derive the Bessel correlation result Eq. (\ref{eq:bess}) cannot be applied to more general objects. For example, there is no known way to relate objects like the fourth-order spatial correlation function
\begin{eqnarray}
\label{eq:R4}
R_{n}(\vec{r}_{1},\vec{r}_{2},\vec{r}_{3},\vec{r}_{4}) = \frac{1}{A(S)} && \times \\
\int_{S}\psi_{n}(\vec{r}_{1}+\vec{q})\psi_{n}(\vec{r}_{2}+\vec{q}) && \psi_{n}(\vec{r}_{3}+\vec{q})\psi_{n}(\vec{r}_{4}+\vec{q})d\vec{q}  \nonumber   
\end{eqnarray}
to the Bessel two-point correlation. The reason lies deep in the very construction of the semiclassical approximation to the quantum propagator, and it is finally connected with the wrong analytical structure of its semiclassical approximation. This interesting subject is beyond the scope of this contribution and we recommend, e.g,  \cite{fredh} for further information.

After mentioning that the semiclassical techniques cannot be directly used to calculate higher order correlations, we are ready to formulate Berry's conjecture. Berry's insight went far beyond the two point case to say that {\it all the statistical properties of the eigenfunction's spatial fluctuations in chaotic systems are described by a superposition of waves with fixed wavenumber and random phases}. In this form the Random Wave Model in principle allows for the calculation of not only the two-point correlation function but any statistical property of irregular eigenfunctions, as we show now. 

\subsection{Random Wave Model as a Gaussian Random Field}
\label{sec:3}
In principle, in order to calculate higher order correlation functions using the RWM, we must insert our definition (\ref{eq:RWM}) into the expression to be averaged (for example  Eq. (\ref{eq:R4})), then we evaluate the average over the random phases and consider the limit $L\to \infty$. This procedure can be of use for very simple averages, but it becomes rapidly complicated with the order of the statistical measure. 

In general, for a $M$-point statistical measure 
\begin{equation}
F(\psi_{n}^{RWM}(\vec{r}_{1};\delta),\ldots,\psi_{n}^{RWM}(\vec{r}_{M};\delta))
\end{equation}
depending on the values of the random wave at fixed points $\vec{r}_{1},\ldots,\vec{r}_{M}$, we must evaluate the integral over the random phases
\begin{eqnarray}
\label{eq:FRWM}
\langle F \rangle&=&\int F(\psi_{n}^{RWM}(\vec{r}_{1};\delta),\ldots,\psi_{n}^{RWM}(\vec{r}_{M};\delta))d \delta \nonumber \\
&=& \int F(\vec{v})P(\vec{v})d\vec{v},
\end{eqnarray} 
where $d\delta=d\delta_{1}\ldots d\delta_{L}$. We also introduced the vector $\vec{v}=(v_{1},\ldots,v_{M})$ and its joint probability distribution 
\begin{equation}
P(\vec{v})=\lim_{N\to \infty} \int \left[\prod_{i=1}^{M}\delta(v_{i}-\psi_{n}^{RWM}(\vec{r}_{i};\delta))\right]d \delta,
\end{equation}
in general a very hard calculation.

Luckily, explicit calculation of the distribution $P(\vec{v})$ for eigenfunction values is not necessary. Due to a straightforward application of the central limit theorem \cite{sto1}, the distribution of eigenfunction values is Gaussian, namely,
\begin{equation}
\label{eq:Pv}
P(\vec{v})=\frac{1}{\sqrt{(2 \pi)^{M}{\rm det\ }{\bf J}}}{\rm e}^{-\frac{1}{2}\vec{v}.{\bf J}^{-1}.\vec{v}},
\end{equation}
where the elements of the so-called correlation matrix ${\bf J}$ (the only set of parameters defining the distribution) are given by
\begin{equation}
\label{eq:Rij}
J_{i,j}=\frac{1}{A(\Omega)}J_{0}(k_{n}|\vec{r}_{i}-\vec{r}_{j}|).
\end{equation}
Equations (\ref{eq:Pv}) and (\ref{eq:Rij}) represent the second (conjectured) ingredient making Berry's conjecture such a powerful tool, and another source of universal behavior in the statistical description of chaotic eigenfunctions. Not only the two-point correlation function is universal, also the distribution of eigenfunction's values is universally described by a Gaussian distribution uniquely fixed by the universal two-point correlation function. The third version of the RWM conjecture now says that {\it eigenfunctions in quantum systems with chaotic classical dynamics have the same statistical properties as a Gaussian random field with a universal correlation function}. This is the form of the conjecture adopted in more recent works \cite{sied1}.

We have seen during this meeting several examples of the robustness and power of this conjecture. Not only the two-point correlation function has been checked, but also the impressive agreement of the RWM with very complicated functionals of the eigenfunctions such as vortex correlations and nodal densities shows the universality of the eigenfunction distributions. 

However, as refined calculations near boundaries were required (for example in \cite{berr2}), two major issues moved into focus. Both of them demanding changes in the requierements to the RWM, and both of them producing deviations from universality and/or Gaussian statistics. In the rest of our contribution we will study these modifications.

\section{Including boundary conditions into the Random Wave Model}
\label{sec:5}
Our first concern to describe realistic eigenfunctions beyond the universal isotropic results of the RWM is the inclusion of boundaries into the model. Inasmuch as the RWM is based on two concepts, namely, a universal two-point correlation and an universal probability distribution, one has to understand how the boundaries could affect each of these ingredients.

This question was posed already in the frame of the statistical description of wavefunctions in disordered systems, and the answer appeared to be quite drastic: the presence of boundaries is not compatible with Gaussian statistics \cite{mirl2}.

This result has delicate consequences, particularly since much of the power and usefulness of the RWM stems from the Gaussian statistics which came with a set of powerful rules to perform calculations (most of all Wick's theorem). Is it possible to keep at least by now a Gaussian theory but to introduce boundary effects in a less destructive way? Contrary to the expectations imported from the theory of disordered systems, the answer to this question is positive, but we need to introduce more carefully the kind of averages our new approach can handle. 

We have seen in section \ref{sec:2} that the original RWM dealt with spatial averages. It is quite clear that spatial averages will in general destroy any information about the boundary, unless they are done in a very special way. For example, if the spatial region $S$ we use to perform the integrations in Eq. (\ref{eq:Rn}) is circular, we will lose any information on the direction where the boundary is located. In fact we will reinforce the Bessel result, as shown in \cite{KJD2}. 

We conclude that any kind of well defined directional or non isotropic effect is visible only when the average is done in such way that no spatial integration is required. In the case of disordered systems this is done by definition, since the different realizations of the disordered potential provide a natural ensemble, but in the case of a single system (as it should be for clean systems) there is no such mechanism.

The only option left is to no longer deal with a single eigenfunction and perform the averages for fixed sets of positions, without any spatial integration and varying the energy index $n$. As we will see, this idea will open the door for a very compact and appealing formulation of Berry's conjecture.

\subsection{Eigenfunction's averages over spectral windows}
\label{sec:6} 
Consider now a functional $F[\psi]$ depending on the set of $M$ positions $\vec{r}_{1},\ldots,\vec{r}_{M}$ as parameters through the values of the function at those points
\begin{equation}
F[\psi]=F(\psi(\vec{r}_{1}),\ldots,\psi(\vec{r}_{M})).
\end{equation}
When we plug different eigenfunctions into $F$, we obtain a fluctuating quantity. In a natural way we define the spectral average of the functional $F$ around energy $e$, indicated by calligraphic symbols from now on, as \cite{KJD1}
\begin{equation}
\label{eq:Fav}
{\cal F}=\frac{1}{\rho_{w}(e)}\sum_{n}w(e-e_{n})F[\psi_{n}]
\end{equation}
where $w(x)$ is a normalized window function around $x=0$ and $\rho_{w}(e)$ is the density of states smoothed over a window $w$. Our objective is to construct a random wave function to describe such averages, inasmuch as Berry's RWM describes spatial averages, and based on similar statistical assumptions.

First we note that our spectral average indeed respects boundary conditions. For example, if one of the positions in the set $\vec{r}_{1},\ldots,\vec{r}_{M}$ lies exactly at the boundary, then all the eigenfunctions must vanish there exactly, not only on average but sharply. How can we implement such hard constraint into the otherwise very smooth Gaussian distribution? The answer is actually quite simple: do not change anything.

\subsection{An ensemble of random waves satisfying arbitrary boundary conditions}
\label{section:7} 

Consider the exact two-point correlation function associated with the spectral average in Eq. (\ref{eq:Fav}),
\begin{equation}
\label{eq:Rex}
R_{w}(\vec{r}_{1},\vec{r}_{2},e)=\frac{1}{\rho_{w}(e)}\sum_{n}w(e-e_{n})\psi_{n}(\vec{r}_{1})\psi_{n}(\vec{r}_{2}),
\end{equation}
and note that it satisfies exactly the boundary conditions. Now, let us construct the following Gaussian distribution
\begin{equation}
\label{eq:PR}
P_{R}(\vec{v})=\frac{1}{\sqrt{(2 \pi)^{M}{\rm det \ }{\bf R}}}{\rm e}^{-\frac{1}{2}\vec{v}.{\bf R}^{-1}.\vec{v}},
\end{equation}
fixed by the system-dependent correlation matrix
\begin{equation}
\label{eq:Rw}
R_{i,j}=R_{w}(\vec{r}_{i},\vec{r}_{j},e),
\end{equation}
depending on the set of positions $\vec{r}_{1},\ldots,\vec{r}_{M}$ and the energy $e$ as parameters.

It is not difficult to check that the probability distribution $P_{R}(\vec{v})$ converges to a delta function $\delta(v_{i})$ when the point $\vec{r}_{i}$ lies at the boundary. Namely, Eqs. (\ref{eq:PR}, \ref{eq:Rw}) define an ensemble of random waves satisfying sharply the boundary conditions. We mention that, although we considered the Dirichlet case for simplicity, any kind of Hermitian boundary condition can be treated exactly in the same way.

Using our construction, we can present yet another version of Berry's conjecture, now with the extra gain including arbitrary boundary conditions: {\it spectral averages of functionals in classically chaotic quantum systems are universally described by a Gaussian distribution fixed by the system-dependent two point correlation function}.

At this point, a natural question arises, namely, what is the connection between the universal predictions of the RWM and the system-dependent predictions of our new ensemble? Luckily, they are compatible, since standard techniques based on the semiclassical expansion of the propagator can be used to calculate the correlation function, and $R_{w}(\vec{r}_{i},\vec{r}_{j},e)$ can be represented in a multiple reflection expansion of the form \cite{sied2,KJD2}
\begin{eqnarray}
R_{w}(\vec{r}_{i},\vec{r}_{j},e)&=&\frac{1}{A(\Omega)}J_{0}(k(e)|\vec{r}_{i}-\vec{r}_{j}|) \\ &+& {\rm \ sum \ over \ reflections \ at \ the \ boundary} \nonumber.
\end{eqnarray}
Therefore a similar expansion of arbitrary averages around the universal RWM results can be constructed. This method was used in \cite{KJD1} to formally show the equivalence between the theory of eigenfunction statistics in disordered metals and the improved form of Berry's conjecture. In particular, all the results of the constraint Random Wave Models proposed to incorporate simple geometries (as in \cite{berr3}) are recovered, extended and generalized using the random wave model with boundaries \cite{KJD0}.

For example, using the Gaussian distribution Eq. (\ref{eq:PR}) together with the following (``one bounce'') two-point correlation function
\begin{eqnarray}
\label{eq:R1b}
R_{1.b}(\vec{r}_{i},\vec{r}_{j},e)=\frac{1}{A(\Omega)}&&J_{0}(k(e)|\vec{r}_{i}-\vec{r}_{j}|)\\+\left|2 \kappa\frac{L_{i}L_{j}\sec{\theta}}{L_{i}+L_{j}}-1\right|^{-1/2}&\times&\frac{\cos\left[k(e)(L_{i}+L_{j})-\phi\right]}{A(\Omega)\sqrt{2 \pi (L_{i}+L_{j})}} \nonumber
\end{eqnarray}
representing the effect of a single reflection with the boundary at point $\vec{r}$ with incidence/reflection angle $\theta$, we can explicitly calculate the effect of such boundary on any statistical measure \cite{KJD2}. Here the local curvature at $\vec{r}$ is $\kappa$ and $L_{i(j)}=|\vec{r}_{i(j)}-\vec{r}|$. The extra phase $\phi$ is related to the topology of the local classical flux around the reflected path and can be also explicitly constructed \cite{brack}.
\begin{center}
\begin{figure}

\resizebox{0.85\columnwidth}{!}{%
  \includegraphics{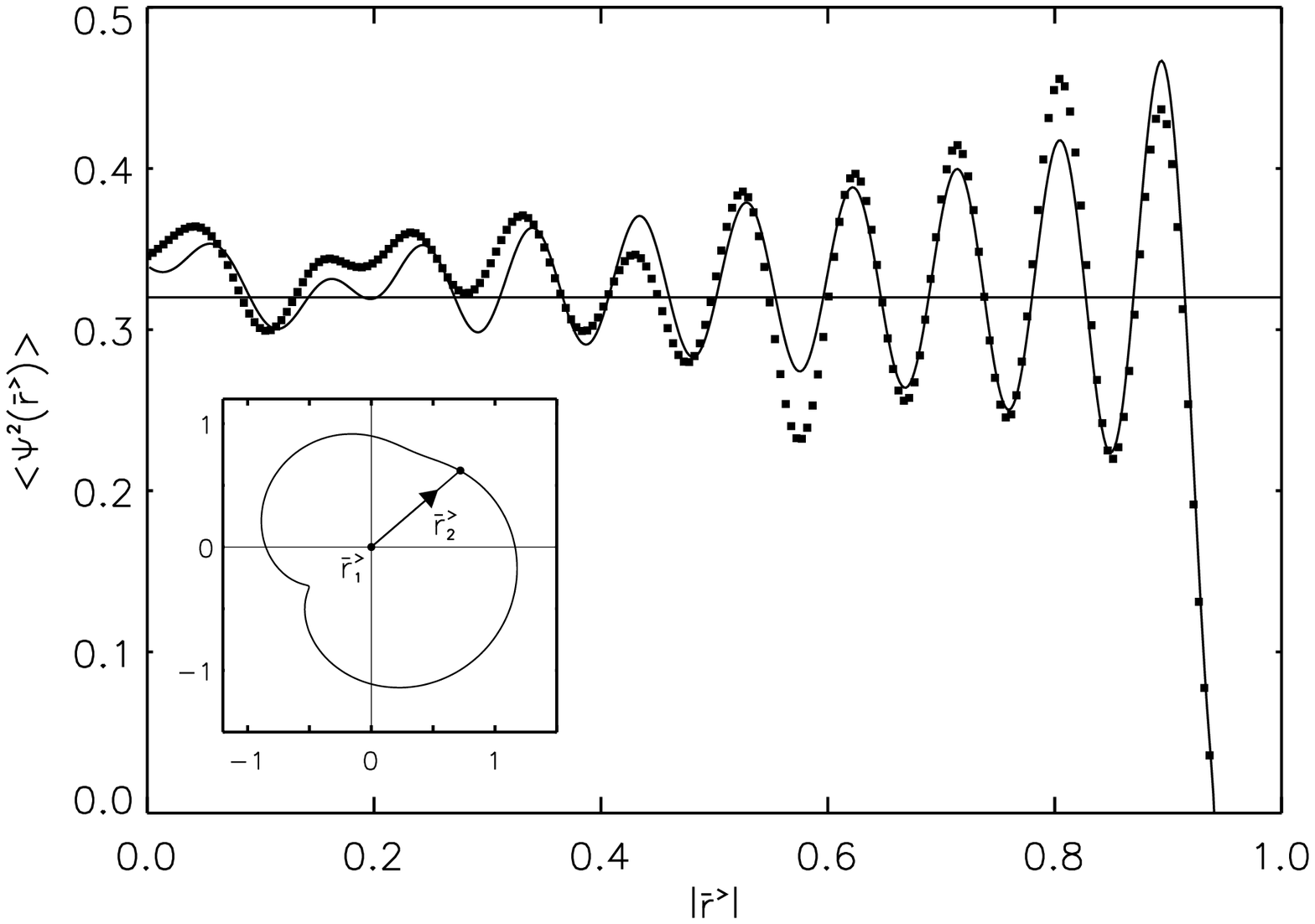} }

\vspace{0.5cm}

\resizebox{0.85\columnwidth}{!}{%
  \includegraphics{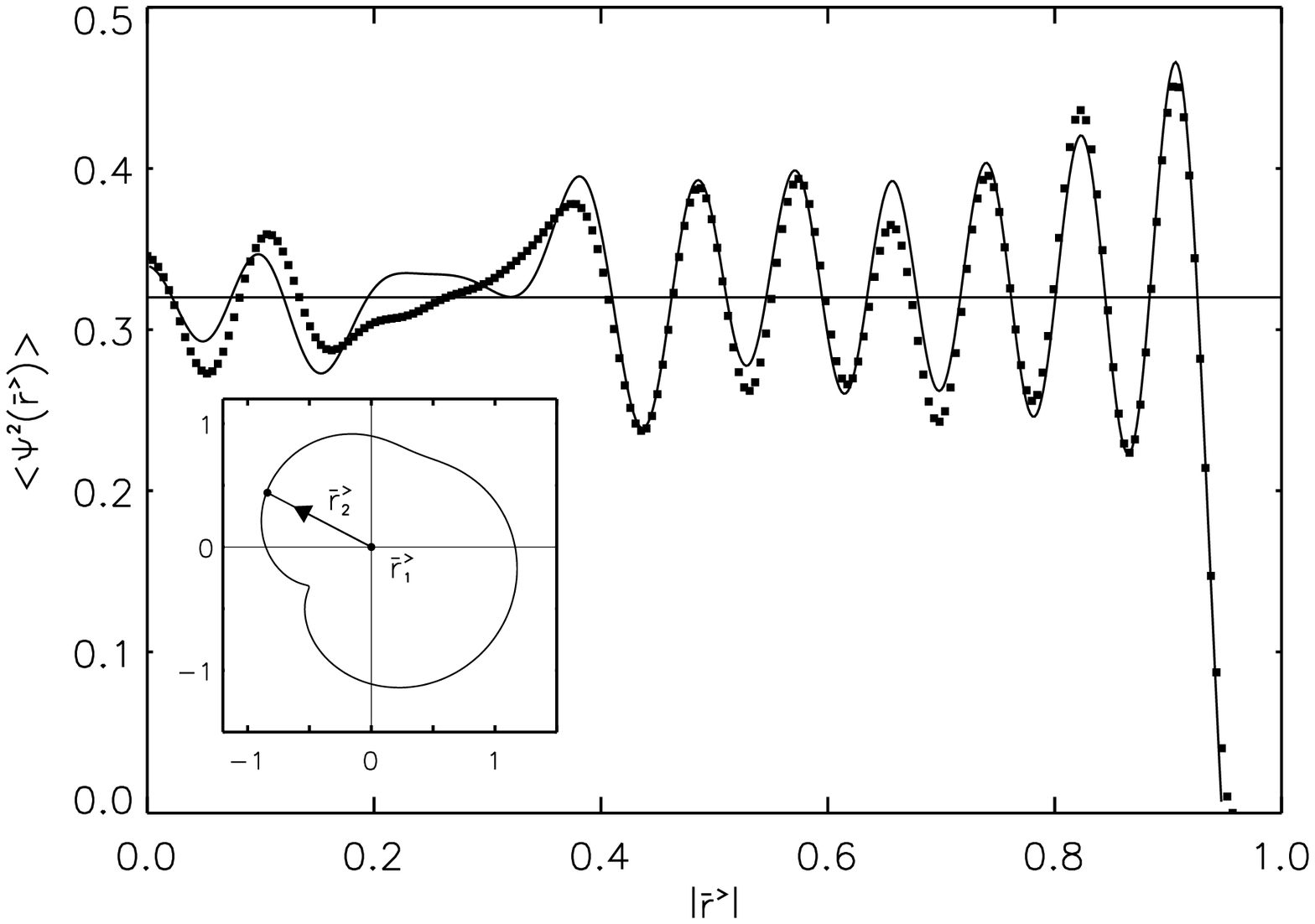} }

\caption{Exact calculation (symbols), universal result (horizontal line) and one-bounce approximation (continuous line) of the average intensity $\langle \psi^{2}(\vec{r}_{2})\rangle=R_{w}(\vec{r}_{2},\vec{r}_{2},e)$ in the cubic conformal billiard \cite{rob}. The average intensity is plotted as a function of the distance $|\vec{r}|=|\vec{r}_{2}-\vec{r}_{1}|$ to the origin $\vec{r}_{1}$ while moving the observation point $\vec{r}_{2}$ along the line indicated in the inset. Non universal effects are clearly visible and well described by the one-bounce correlation function of Eq. (\ref{eq:R1b}) (diffraction effects must be incorporated). Note in particular how the boundary condition (Dirichlet) is respected}
\label{fig:1}    
\end{figure}
\end{center}
Figure (\ref{fig:1}) shows the average intensity $\langle \psi^{2}(\vec{r}_{2})\rangle$ (given by $R_{w}(\vec{r}_{2},\vec{r}_{2},e)$ in Eq. (\ref{eq:Rex})) as a function of the distance $|\vec{r}|=|\vec{r}_{2}-\vec{r}_{1}|$ from the origin $\vec{r}_{1}$ to the observation point $\vec{r}_{2}$ as it moves towards the boundary of a particular chaotic billiard, shown in the inset. Symbols are exact numerical results obtained from the numerically exact eigenfunctions, while the continuous line is given by $R_{1.b}(\vec{r}_{2},\vec{r}_{2},e)$ (including diffraction effects). As it can be seen, the presence of the boundary drastically modifies the local structure of the averages when compared with the $1/A(\Omega)=1/ \pi$ constant result. 

Our ensemble of random waves with boundaries used our freedom to incorporate non-universal effects by means of the two-point correlation function, leaving as the only conjecture a statement about the universal Gaussian fluctuations of irregular eigenfunctions. However, to construct an ensemble respecting all the properties of the exact eigenfunctions, every member of the ensemble should be properly normalized. As we will see, Gaussian statistics is not compatible with this constraint, and the Gaussian conjecture must be modified.

\section{Projecting the ensemble of random waves onto the set of normalized functions}
\label{sec:8}

Our goal in this section is to construct an ensemble of random functions, respecting sharply the boundary conditions, such that every member of the ensemble is exactly normalized. This is more than an academic question: it is fundamental to have normalization correctly taken into account, because otherwise the results of the ensemble average will contain spurious fluctuations affecting any statistic beyond $M=2$, as we shall explain. 

The so-called normalization problem was first noted in \cite{mirl1} where the corrections due to normalization for disordered metals were found. In the last section of this contribution, we will show how to obtain such corrections in the appropriate limit of our re-formulated Random Wave Model, giving strong support to our construction. 

In order to appreciate the kind of spurious effects which we get using Gaussian statistics, we consider the functional 
\begin{equation}
F[\psi]=\int \psi^{2}(\vec{r})d\vec{r},
\end{equation}
which should have zero variance if every member of the ensemble is properly normalized, since we have $F[\psi]=1$ over the set of normalized eigenfunctions. However, let us calculate explicitly this variance by using our random wave model:
\begin{eqnarray}
\label{eq:NP}
\langle F^{2} \rangle-\langle F\rangle ^{2}&=&\int\langle \psi^{2}(\vec{r})\psi^{2}(\vec{r}')\rangle d\vec{r}d\vec{r}'-1 \\
&=&2\int R_{w}^{2}(\vec{r},\vec{r}',e)d\vec{r}d\vec{r}', \nonumber
\end{eqnarray}
where we have used the identity $\int R_{w}(\vec{r},\vec{r},e)d\vec{r}=1$. Result Eq. (\ref{eq:NP}) is obviously not zero, and this indicates that if we calculate the spectral average of a functional of order higher than two, fluctuations of the normalization integral will produce spurious contributions. 

The mathematical problem is the search for a probability distribution over the set of normalized wavefunctions, as close as possible to the Gaussian (since we have very strong evidence for it), and uniquely fixed by the given two-point correlation function (since this object is calculated exactly and does not suffer from normalization problems). Of course ``as close as possible'' should be defined properly, but presently the structure and topology of the space of distributions living on the set of normalized functions and obeying a fixed two-point correlator is still poorly understood; so following an educated guess, we will try to construct one of such distributions hoping that its predictions match the results for disordered systems. However, before carrying out this program, we need to take a further step.

\subsection{The non-local and non-stationary character of the normalization constraint}
\label{section:9} 

One of the convenient properties of the Gaussian distribution Eq. (\ref{eq:PR}) is its {\it local} structure. Locality in this context means the following: let us define 
\begin{eqnarray}
\vec{v}_{N}&=&(\psi(\vec{r}_{1}),\ldots,\psi(\vec{r}_{N})) \nonumber \\
\vec{v}_{M}&=&(\psi(\vec{r}_{1}),\ldots,\psi(\vec{r}_{N}),\ldots,\psi(\vec{r}_{M})) \nonumber 
\end{eqnarray}
with $N < M$, and consider a functional $G[\psi]$ depending on the subset of positions $\vec{r}_{1},\ldots,\vec{r}_{N}$. It is then true that its Gaussian average ${\cal G}$ satisfies
\begin{eqnarray}
{\cal G} &=&\frac{1}{\sqrt{(2 \pi)^{M}{\rm det\ }{\bf R}}}\int G(\vec{v}_{N}){\rm e}^{-\frac{1}{2}\vec{v}_{M}.{\bf R}^{-1}.\vec{v}_{M}}d\vec{v}_{M} \\ \nonumber
&=& \frac{1}{\sqrt{(2 \pi)^{N}{\rm det\ }{\bf R}'}}\int G(\vec{v}_{N}){\rm e}^{-\frac{1}{2}\vec{v}_{N}.{\bf R}'^{-1}.\vec{v}_{N}}d\vec{v}_{N},
\end{eqnarray}
where ${\bf R}'$ is a correlation matrix obtained by simply removing the last $M-N$ indeces from ${\bf R}$ (from now on we suppress the $w$ index of the correlation matrix). In other words, the average of a functional depending on a given subset of positions in still Gaussian (stationarity) and depends only on the eigenfunction correlations among only those points (locality).

The locality and stationarity of the Gaussian distributions are at the very heart of its generality. They mean that the distribution has the same form independent of how many points the functional has and that the averages depend only on the points where the functional is defined. Because of this, the most general form of the Gaussian probability distribution is the continuous limit of Eq. (\ref{eq:PR}), that is, a functional distribution
\begin{eqnarray}
\label{eq:PRinf}
P_{R}[\psi]{\cal D}[\psi]&=&\lim_{M->\infty}\frac{1}{\sqrt{(2 \pi)^{M}{\rm det\ }{\bf R}}} \\ &\times &{\rm e}^{-\frac{1}{2}\sum_{i,j}^{M}\psi(\vec{r}_{i})\left({\bf R}^{-1}\right)_{i,j}\psi(\vec{r}_{j})}d\psi(\vec{r}_{1})\ldots d\psi(\vec{r}_{M}) \nonumber \\
&=&\frac{{\cal N}_{M}}{\sqrt{(2 \pi)^{M}{\rm det\ }\hat{R}}}{\rm e}^{-\frac{1}{2}\langle\psi|\hat{R}^{-1}|\psi \rangle}{\cal D}[\psi] \nonumber
\end{eqnarray}
where, after proper scaling of $\psi(\vec{r}_{i})$ with $M$ absorbed by ${\cal N}_{M}$, the quadratic form in the exponent is
\begin{eqnarray}
\label{eq:cuad}
\langle\psi|\hat{R}^{-1}|\psi \rangle&=&\lim_{M->\infty}\frac{1}{M}\sum_{i,j}^{M}\psi(\vec{r}_{i})({\bf R}^{-1})_{i,j}\psi(\vec{r}_{j}) \nonumber \\
&=& \int \psi(\vec{r})({\bf R}^{-1})(\vec{r},\vec{r}')\psi(\vec{r}')d\vec{r}d\vec{r}',
\end{eqnarray}
while $\hat{R}$ (referred from now on as the ``covariance'' or ``correlation'' operator) with matrix elements $\left({\bf R}\right)_{i,j}=\langle \vec{r}_{i}|\hat{R}|\vec{r}_{j}\rangle$ is naturally defined through the spectral window function $w(x)$ and the system Hamiltonian $\hat{H}$ as
\begin{equation}
\label{eq:Rop}
\hat{R}=\frac{w(e-\hat{H})}{{\rm Tr\ }w(e-\hat{H})}.
\end{equation}
The notations $[\hat{O}]_{i,j}$, $(\hat{O})(\vec{r}_{i},\vec{r}_{j})$ and $\langle \vec{r}_{i}|\hat{O}|\vec{r}_{j}\rangle$ will be used from now on to indicate matrix elements.

Our expressions (\ref{eq:PRinf},\ref{eq:cuad}) and (\ref{eq:Rop}) have the very important property of being independent of the representation, in the sense that they define the probability of finding a given function $\psi$, not a particular set of values for their projection along a particular basis.

Now let us consider the effect of normalization constraints on the probability distribution in Eq. (\ref{eq:PRinf}). The requirement that all the members of our ensemble are normalized necessarily implies that the correct distribution that we indicate with $P_{R}^{GAP}$ (anticipating its ``Gaussian-Projected'' character) must be sharply concentrated on the unit sphere in the space of functions,
\begin{equation}
\label{eq:prim}
P_{R}^{GAP}[\psi]{\cal D}[\psi]\sim \delta(\langle \psi |\psi\rangle-1){\cal D}[\psi].
\end{equation}
From this condition we see that, whatever we do, the normalization constraint destroys the stationary and local nature of the distribution, since any set of points is connected with any other through the normalization. In other words, and in sharp contrast to the Gaussian distribution, averages of functionals defined over any subset of positions will feel the fluctuations of the eigenfunctions at any point of the space. The ensemble of normalized random functions cannot be neither local, nor stationary. 

However, evidence shows that a Gaussian (local and stationary) distribution must be a very good approximation, and we will see that we can indeed construct a distribution over the unit sphere as close to the Gaussian as we wish. Interestingly, although we start talking about a single particle inside a quantum billiard, the solution will come from the field of quantum statistical physics. We mention that another intersection between random waves and quantum statistical physics is presented in \cite{hell}.

\subsection{The connection with quantum statistical physics}
\label{section:10} 

Besides being defined over the unit sphere in the space of functions, our distribution must fulfill the basic requirement that the associated two-point correlation function is known:
\begin{equation}
\label{eq:seg}
\int \psi(\vec{r}_{i})\psi(\vec{r}_{j})P_{R}^{GAP}[\psi]{\cal D}[\psi]=\langle \vec{r}_{i}|\hat{R}|\vec{r}_{j}\rangle.
\end{equation} 
Equations (\ref{eq:prim},\ref{eq:seg}) define our problem. Not much is known about the possible solution, but at least it is easy to see that it is not unique \cite{gold}. At this point our hope is that the physics of irregular eigenfunctions can help with extra conditions when looking for a unique solution, but this issue remains open.

Surprisingly, exactly the same question has had a long history in the field of quantum statistical physics (we follow \cite{gold}). In this context, one is interested on the possible distribution $P_{\rho}[\psi]$ of individual {\it normalized} wavefunctions for systems in thermal equilibrium characterized by a given density operator 
\begin{equation}
\hat{\rho}=\frac{{\rm e}^{-\beta \hat{H}}}{{\rm Tr\ } {\rm e}^{-\beta \hat{H}}},
\end{equation}
where $\hat{H}$ is the Hamiltonian and $\beta$ the inverse temperature. Consistency with the results of quantum statistical physics demands that averages of expectation values $\langle \psi |\hat{O}|\psi\rangle$, weighted by $P_{\rho}[\psi]$, should give the usual result 
\begin{equation}
\int\langle \psi |\hat{O}|\psi\rangle P_{\rho}[\psi] {\cal D}[\psi]={\rm Tr\ }(\hat{\rho} \hat{O}),
\end{equation}
for arbitrary $\hat{O}$. This in turn implies
\begin{equation}
\label{eq:ter}
\int \psi(\vec{r}_{i})\psi(\vec{r}_{j})P_{\rho}[\psi]{\cal D}[\psi]=\langle \vec{r}_{i}|\hat{\rho}|\vec{r}_{j}\rangle,
\end{equation} 
and then it is clear that $P_{\rho}$ and $P_{R}^{GAP}$ satisfy the same conditions, but in the first case we talk of a ``correlation matrix'' instead of a ``density operator''. In fact, $\hat{R}$ is also a positive, analytical function of the Hamiltonian with unit trace, so {\it it is} a density operator corresponding to the microcanonical distribution. 

One could think, inasmuch as in the case of few degrees of freedom, that we cannot choose among all the possible solutions of conditions (\ref{eq:prim},\ref{eq:seg}). However, this is not the case. There is a guiding principle frequently used and well known in the framework of classical and quantum statistical physics, known as ``Principle of Maximum Entropy'' (PME) saying that, for a given set of constraints, the best option for a probability distribution is the one which maximizes the associated entropy functional. The exact formulation and the implications of this principle go beyond the scope of this paper, and the reader is encouraged to read the excellent literature available \cite{PME}. However, as it is shown in \cite{gold}, when applied to the present problem the PME selects one particular kind of distribution, known as the ``Gaussian Projected Ensemble'' and given by
\begin{equation}
\label{eq:PRgap}
P_{R}^{GAP}[\psi]=\int \delta\left(\psi-\frac{\phi}{\sqrt{\langle \phi|\phi \rangle}}\right)\langle \phi|\phi \rangle P_{R}[\phi]{\cal D}[\phi].
\end{equation}
We mention that in the microcanonical case (our main interest here since it defines the Random wave Model), this is the unique solution to the maximization problem.

Following a long and successful tradition, we assume that the methods and techniques of statistical physics can be applied, or at least can shade some light, to the nature of systems with complex behavior but few degrees of freedom. We then propose the following version of Berry's conjecture, incorporating exactly both, boundary and normalization constraints: {\it In systems with classically chaotic dynamics, spectral averages of functionals defined over the set of eigenfunctions, are given by the corresponding average over the Gaussian Projected Ensemble with fixed system-dependent covariance matrix}. In other words, we conjecture that the statistical fluctuations of irregular eigenfunctions are universally described by the Gaussian Projected Ensemble.

Of course, only the successful application of the conjecture will decide whether it expresses or not a fundamental property of irregular eigenfunctions, and this is work is in progress. There are, however, some predictions that can be checked and contrasted with existing results coming from the theory of disordered systems (a connection which has been very useful in the past) \cite{sto1}. Before that, however, we must be able to operate with the GAP distribution, something that has not been done until now due to its non-local and non-stationary structure.

\subsection{Correlation functions for the GAP ensemble}
In this section we briefly discuss recent work on the correlation functions associated with the GAP measure \cite{JD3}. Some of this results are currently used for specific calculations, hoping that they can be tested against numerical or experimental evidence, and others can be directly compared with known results for statistics of eigenfunctions in disordered systems. To be more precise, we first introduce some basic concepts. 

A local functional over the $N$- dimensional subspace ${\cal H}_{N}$ of the Hilbert space ${\cal H}$ is a function of
the values of the wavefunction at the given (finite) set of $N$ positions  
\begin{equation}
F[\psi]=F(\psi(\vec{r}_{1}),\ldots,\psi(\vec{r}_{N})).
\end{equation}

Using the definition Eq. (\ref{eq:PRgap}), the GAP average of $F$ is given by the functional integral
\begin{eqnarray}
\label{eq:FGAP}
{\cal F}_{GAP}&=&\frac{{\cal N}}{\sqrt{{\rm det}\hat{R}}}\int F\left[\frac{\psi}{|\psi|}\right]
{\rm e}^{-\frac{1}{2}\langle \psi |\hat{R}^{-1}| \psi \rangle}|\psi|^{2}D[\psi] \nonumber \\
&=& \frac{{\cal N}}{\sqrt{{\rm det}\hat{R}}}\int F\left(\frac{\psi(\vec{r}_{1})}{|\psi|},\ldots,\frac{\psi(\vec{r}_{N})}{|\psi|}\right) \nonumber \\
\times & & {\rm e}^{-\frac{1}{2}\langle \psi |\hat{R}^{-1}| \psi \rangle}|\psi|^{2}D[\psi]
\end{eqnarray}
where $<,>$ is the inner product on ${\cal H}$, $|\psi|=|\langle \psi|\psi \rangle |^{1/2}$ the norm
and ${\cal N}$ is an (infinite) normalization constant.
The operator $\hat{R}^{-1}$ (also known as $\hat{\rho}^{-1}$) is well defined because of the properties of the function $w(z)$, which is assumed to be analytic and positive

For future reference, the result of the Gaussian (non-projected) average is given by
\begin{eqnarray}
{\cal F}_{G}&=&\frac{{\cal N}}{\sqrt{{\rm det}\hat{R}}}\int F[\psi]
{\rm e}^{-\frac{1}{2}\langle \psi |\hat{R}^{-1}| \psi \rangle}D[\psi] \nonumber \\
&=&\int F\left(\psi_{1},\ldots,\psi_{N}\right)
\frac{{\rm e}^{-\frac{1}{2}\vec{\psi}.(\hat{R}_{N})^{-1} .\vec{\psi}}}{\sqrt{(2 \pi )^{N}{\rm det_{N}} \hat{R}_{N}}}d\vec{\psi} \\
&=&{\cal F}_{G}(\hat{R}_{N}), \nonumber
\end{eqnarray}
which is calculated by means of the usual $N$-dimensional Gaussian integral over the vector variable $\vec{\psi}=(\psi_{1},\ldots,\psi_{N})$. The 
correlation matrix $\hat{R}_{N}$ is the projection of the operator $\hat{R}$ 
on ${\cal H}_{N}$ with entries given by
\begin{equation}
[\hat{R}_{N}]_{i,j}=\langle \vec{r}_{i}|\hat{R}| \vec{r}_{j} \rangle 
{\rm  \ for \ } \vec{r}_{i},\vec{r}_{j} \in {\cal H}_{N},
\end{equation}
while ${\rm det}_{N}$ is the determinant on ${\cal H}_{N}$.

It is essential that for an arbitrary operator $\hat{A}$ on ${\cal H}$ and an analytical function $f(z)$,
\begin{equation}
\label{eq:care}
f(\hat{A})_{N} \neq f(\hat{A}_{N}),
\end{equation}
although both sides of the equation are defined over the same space ${\cal H}_{N}$. 
In particular,the matrix $(\hat{R}_{N})^{-1}$ is constructed by projecting $\hat{R}$ and then
inverting, rather than projecting $\hat{R}^{-1}$. 

The pure technical problem with the GAP  averages
is that, contrary to the Gaussian case, one cannot directly integrate out the components of $\psi$ belonging
to the subspace ${\cal H}/{\cal H}_{N}$. This problem finally boils down to the appearance of the norm $|\psi|$, 
with components over the whole ${\cal H}$, as a denominator in ${\cal F}_{GAP}$, as it is seen
in Eq. (\ref{eq:FGAP}). However, if $F$ is a {\it homogeneous functional} 
\begin{equation}
F^{(2\mu)}(\lambda \psi(\vec{r}_{1}),\ldots,\lambda \psi(\vec{r}_{N}))=\lambda ^{2 \mu}
F^{(2 \mu)}(\psi(\vec{r}_{1}),\ldots,\psi(\vec{r}_{N}))
\end{equation}
these denominators can be collected and written as an exponential by means of the following identity 
\begin{equation}
\frac{1}{|\psi|^{2\nu}}=\frac{1}{\Gamma(\nu)}\int_{0}^{\infty}x^{\nu-1}{\rm e}^{-x\langle \psi|\psi \rangle}dx
\end{equation}
valid for ${\rm Re }(\nu) >0$, where $\Gamma(x)$ is the gamma function. The integrals in Eq. (\ref{eq:FGAP}) are now purely Gaussian, and carefully considering Eq. (\ref{eq:care})
the desired reduction of the integration over ${\cal H}$ to integration over ${\cal H}_{N}$ is achieved by straightforward
discretization of the functional integral.

The results depend on the order of homogeneity $2 \mu$. For $\mu=0$
 and $\mu=1$ we have
\begin{eqnarray}
{\cal F}_{GAP}^{(0)}&=&{\cal F}_{G}^{(0)}(\hat{R}_{N})
-2 \frac{\partial}{\partial \alpha}{\cal F}_{G}^{(0)}(\hat{R}_{N}+\alpha (\hat{R}^{2})_{N}) |_{\alpha=0}, \nonumber \\
{\cal F}_{GAP}^{(2)}&=&{\cal F}_{G}^{(2)}(\hat{R}_{N}).
\end{eqnarray}
The result for $2 \mu=2$ implies as particular case that both the Gaussian and GAP averages 
give the same result for the associated correlation (density) operator, but it is far more general. For $0 < \mu <1$ and $1< \mu <2$ we have not been able to calculate the corresponding integrals.

For $\mu \geq 2$ we first define a one-parameter deformation of the correlation operator
\begin{equation}
\label{eq:Q}
\hat{Q}(x)=\hat{R}(1+2x \hat{R})^{-1},
\end{equation} 
where the non-negativity of $\hat{R}$ assures existence of the inverse for $x \geq 0$ and renders $\hat{Q}(x)/{\rm Tr }Q(x)$ a density operator as well. With this definition the GAP average is given by the following one-parameter integral 
\begin{equation}
\label{eq:main}
{\cal F}_{GAP}^{(2\mu)}=\frac{1}{\Gamma (\mu-1)}\int _{0}^{\infty} \frac{x^{\mu-2}}{\sqrt{ {\rm det}(1+2 x \hat{R})}}
{\cal F}_G^{(2 \mu)}(\hat{Q}(x)_{N})dx.
\end{equation}
Equation (\ref{eq:main}) together with the definition Eq. (\ref{eq:Q}) is the main result of this analysis. 
It allows us to calculate the GAP averages in terms of the Gaussian ones.  

It is important to remark again that {\it normalization effects are non-local} in the sense that any average respecting
normalization (as the GAP), cannot be defined only in terms of the projected density matrix $\hat{R}_{N}$ acting on ${\cal H}_{N}$. The degrees of freedom associated with the complement ${\cal H}/{\cal H}_{N}$ appear through the terms $(\hat{R}^{2})_{N},(\hat{R}^{3})_{N}, \ldots$ in the Taylor expansion of $Q(x)$. As an example consider, for $|\vec{r}_{i}\rangle,|\vec{r}_{j}\rangle \in {\cal H}_{N}$,
\begin{eqnarray}
\label{eq:care2}
(\hat{R}^{2})(\vec{r}_{i},\vec{r}_{j})&=&\langle \vec{r}_{i}|\hat{R}^{2}|\vec{r}_{j}\rangle \\
&=&\frac{1}{A(\Omega)}\int\langle \vec{r}_{i}|\hat{R}|\vec{r}\rangle\langle \vec{r}|\hat{R}|\vec{r}_{j}\rangle d\vec{r} \nonumber \\
&=&\frac{1}{A(\Omega)}\int R(\vec{r}_{i},\vec{r})R(\vec{r},\vec{r}_{j})d\vec{r}. \nonumber 
\end{eqnarray}  
Eq. (\ref{eq:care2}) shows that any dependence non-linear in $\hat{R}$ will connect the points where the functional is defined with points in the whole space. Since any GAP average (except the $2\mu =2$ case) is a non-linear function of $\hat{R}$, and not of $(\hat{R})_{N}$, non-local effects are a generic feature of this distribution. 

\section{The relation between the Gaussian Projected and the Gaussian ensembles}
\label{section:11}

Now we proceed with the asymptotic expansion of (\ref{eq:main}). The key point is the
analytical structure of the function 
\begin{equation}
p(z)={\rm det}(1+2z\hat{R}).
\end{equation}
We note that $p(z)$ is the
spectral determinant of the operator $\hat{R}$ and then an entire function with zeros at, and only at,
\begin{equation}
z_{j}=-\frac{1}{2R(e_{j})},{\rm \ }e_{j} \in {\rm \ spectrum \ of \ }\hat{H}.
\end{equation}
Since the operator $\hat{R}$ is a non-negative function of the Hamiltonian, $z_{j}$ is real negative for all $j$.  
This indicates that, contrary to $x <0$, for $x \geq 0$ the function $p(x)$ cannot be oscillatory, and the following 
expansion is uniform in $x$
\begin{eqnarray}
\label{eq:exp1}
\frac{1}{\sqrt{ {\rm det}(1+2 x \hat{R})}}&=&{\rm exp\ }\left[-\frac{1}{2} {\rm Tr\ log}(1+2x \hat{R})\right] \\ &=& {\rm e}^{-x} 
{\rm exp\ }\left[\frac{1}{2} \sum_{k=2}^{\infty}\frac{(-2x)^{k}}{k}{\rm Tr\ }\hat{R}^{k}\right]. \nonumber 
\end{eqnarray} 
Eq. (\ref{eq:exp1}) is an expansion of the function 
$1/ \sqrt{ {\rm det}(1+2 x \hat{R})}$ in powers of $x$, well behaved upon integration.

The operation that makes our results divergent is the following
expansion of the deformed density matrix
\begin{equation}
\label{eq:exp2}
\left(\frac{\hat{R}}{1+2x\hat{R}}\right)_{N}=\sum_{k=0}^{\infty}(-2x)^{k}(\hat{R}^{k+1})_{N},
\end{equation} 
since its radius of convergence demands $2x < R(e_{1})^{-1}$ while the integration demands $x \to \infty$,
rendering the expansion of the full integral just asymptotic.

Equations (\ref{eq:exp1}) and (\ref{eq:exp2}) provide a systematic expansion of the integrand in Eq. (\ref{eq:main})
in powers of $x$. After term-by-term (trivial) integration, we get the desired asymptotic
expansion of the GAP average around the Gaussian result. It is then of importance to identify the small parameters of this expansion, in order to have control over the successive approximations for specific situations.

\subsection{Physical interpretation of the small parameters}
\label{sec:12} 

In order to identify the small parameter of the expansion, we remark again that $\hat{R}$ is adensity operator and then it is non-negative and of unit trace 
${\rm Tr\ }\hat{R}=1$. Then 
\begin{equation}
 {\rm Tr\ }\hat{R}^{n}<{\rm Tr\ }\hat{R}^{m}, {\rm \ for \ }n>m,
\end{equation}
holds, and the asymptotic expansion can be organized in ascending
powers of the covariance (density) operator, namely, the order of a term is given by the total power of $\hat{R}$. 
These considerations  do not hold for the projected density matrix: $\hat{R}_{N}$
can appear to any power without affecting the order of the corresponding term in the expansion. For example,
 given $|\vec{r}_{i} \rangle,|\vec{r}_{j} \rangle \in {\cal H}_{N}$,
\begin{eqnarray}
\left[\hat{R}^{2}\right]_{i,j} && {\rm \  is \ of \ second \ order,\  but} \nonumber \\
\left[\hat{R}\right]_{i,j} \left[\hat{R}\right]_{i,j} && {\rm \  is \ still \ of \ first \ order.} \nonumber
\end{eqnarray}

In order to interpret physically the small parameter, we consider two examples. 
\begin{itemize}
\item For the microcanonical case of main interest for us, 
the function $w(z)$ in (\ref{eq:Rop}) is a window function of width $\epsilon$. It is easy to show that ${\rm Tr\ }\hat{R}^{2}$ 
is proportional to $\Delta / \epsilon$ where $\Delta$ is the mean level spacing. Then we see that 
{\it for the microcanonical ensemble the small parameter of the asymptotic expansion is $1/L$ where $L$
is the number of energy levels inside the energy window}.
\item For the canonical ensemble at inverse temperature $\beta=1/k_{B}T$ the covariance operator reads $w(z)={\rm e}^{-\beta z}$, and a simple calculation
gives ${\rm Tr}\hat{R}^{2} \sim \Delta/k_{B}T$, so {\it for the canonical ensemble, the small parameter is the ratio
between the mean level spacing and the thermal energy}.
\end{itemize}

Having now control over the asymptotic expansion, we check whether its zero-order term actually gives the Gaussian results.

\subsection{Recovering the Gaussian limit}
\label{section:13}

From the theoretical point of view, the most important property of the asymptotic expansion is that it provides a way
to calculate perturbatively the GAP averages starting with the Gaussian case and keeping the approximation under control.
To see how we obtain to first order the Gaussian average, we truncate both expansions (\ref{eq:exp1}) and (\ref{eq:exp2})
to first order in $\hat{R}$ to obtain
\begin{eqnarray}
\frac{1}{\sqrt{ {\rm det}(1+2 x \hat{R})}} \simeq {\rm e}^{-x} +O({\rm Tr\ }\hat{R}^{2}),\\
\left[\frac{\hat{R}}{1+2x\hat{R}}\right]_{N} \simeq \hat{R}_{N} +O(\hat{R}^{2}).\nonumber
\end{eqnarray} 
Substitution in Eq. (\ref{eq:main}) gives
\begin{eqnarray}
\label{eq:GGAP}
{\cal F}_{GAP}^{(2\mu)}&=&\frac{1}{\Gamma (\mu-1)}\int _{0}^{\infty}x^{\mu-2}{\rm e}^{-x}
{\cal F}_G^{(2 \mu)}(\hat{R}_{N})dx +O((\hat{R}^{2})_{N})\nonumber \\
&=&{\cal F}_G^{(2 \mu)}(\hat{R}_{N})+O((\hat{R}^{2})_{N}).
\end{eqnarray}
Where the definition of the gamma function was used. Our result Eq. (\ref{eq:GGAP}) shows that for ${\rm Tr\ }\hat{R}^{2} \to 0$ (which means high number of levels inside the window function in the microcanonical case or high temperature limit in the canonical), the 
GAP and Gaussian averages coincide. Although it looks intuitively trivial, 
it is of importance since it is obtained keeping the approximation under strict control.

All results based on the asymptotic expansion have a limited accuracy. Poincar\'e criteria for asymptotic
series demand those to be truncated just when the typical term stops decreasing with the order. It is easy to see that,
the higher the order $2\mu$ of the functional, the earlier one must truncate the expansion. The method fails if not even the very first term can be used. This happens for functionals of certain order $2\mu$ inversely related to
${\rm Tr\ }\hat{R}^{2}$. This actually means that the most interesting effects, those beyond a perturbation expansion around
the Gaussian result, occur for higher correlation functions.

However, already for small-order correlations truncated at the first term of the asymptotic expansion, we obtain interesting results, as we will see now. 

\section{An example: asymptotic expansion of the intensity correlator}
\label{section:14}

As a non-trivial example of the method, we calculate the first correction due to normalization to the 
intensity-intensity correlation function, defined by the local homogeneous functional 
\begin{equation}
Y[\psi]=\psi^{2}(\vec{r}_{i})\psi^{2}(\vec{r}_{j}).
\end{equation}
In this case $N=2$, the subspace of interest ${\cal H}_{N}= {\cal H}_{2}$ is expanded by the states $\{ |\vec{r}_{i} \rangle,|\vec{r}_{j} \rangle \}$ and $2\mu=4$. 

In order to apply our ideas, we first identify the Gaussian result ${\cal Y}_{G}(\hat{Q}(x)_{N})$ in terms of the deformed density operator $\hat{Q}(x)_{N}$. 
This is a standard result founded by using the Wick contractions as
\begin{eqnarray}
{\cal Y}_{G}(\hat{Q}(x)_{2})&=&[\hat{Q}(x)]_{i,i}[\hat{Q}(x)]_{j,j}+2[\hat{Q}(x)]_{i,j}[\hat{Q}(x)]_{i,j} \nonumber \\
&=&\left[\frac{\hat{R}}{1+2x \hat{R}}\right]_{i,i}\left[\frac{\hat{R}}{1+2x \hat{R}}\right]_{j,j} \nonumber \\
&+&2\left[\frac{\hat{R}}{1+2x \hat{R}}\right]_{i,j}\left[\frac{\hat{R}}{1+2x \hat{R}}\right]_{i,j}.
\end{eqnarray}

Next, following the idea of an asymptotic expansion in terms of $1/L$ with $L$ the number of energy levels inside the energy window, we use the expansions (\ref{eq:exp1}) and (\ref{eq:exp2}) up to $\hat{R}^{2}\sim 1/L$. We get, after substitution in Eq. (\ref{eq:main}) and calculating
the elementary integrals,
\begin{eqnarray}
\label{eq:Y}
{\cal Y}_{GAP}&=&{\cal Y}_{G}(\hat{R}_{2}) \\
&-&2([\hat{R}^{2}]_{i,i}[\hat{R}]_{j,j}+[\hat{R}^{2}]_{j,j}[\hat{R}]_{i,i}-[\hat{R}]_{i,i}[\hat{R}]_{j,j}{\rm Tr\ }\hat{R}^{2}) \nonumber \\  &+& O(\hat{R}^{3}), \nonumber
\end{eqnarray}
where ${\cal Y}_{G}(\hat{R}_{2})$ is the usual Gaussian result
\begin{equation}
{\cal Y}_{G}(\hat{R}_{2})=[\hat{R}]_{i,i}[\hat{R}]_{j,j}+2[\hat{R}]_{i,j}[\hat{R}]_{i,j}.
\end{equation}
It is easy to see that the extra terms in Eq. (\ref{eq:Y}) render the variance of the normalization integral Eq. (\ref{eq:NP}) exactly zero. This fact appear more clearly when we consider the universal limit of the theory, as we do next. 

\subsection{Universal limit of the intensity correlator: connection with the theory of disordered systems}
\label{section:15}

Our results so far are the asymptotic expansions of the exact GAP averages up to second order in the small parameter. We must remember, however, that for the microcanonical case our results can be further expanded around the universal covariance matrix with entries defined through the Bessel function. The interplay of these two expansions is a delicate matter, but at least we can see what happens when we restrict ourselves to the isotropic result for the correlation function.

In order to find the universal limit of our result Eq. (\ref{eq:Y}), we use the Bessel form of the correlation function Eq. (\ref{eq:bess}), the expression Eq. (\ref{eq:care2}) for the matrix elements $[\hat{R}^{2}]_{i,i} =(\hat{R}^{2})(\vec{r}_{i},\vec{r}_{i})$ and the continuum version of the trace to obtain:
\begin{eqnarray}
\label{eq:Yun}
A(\Omega)^{2}{\cal Y}_{GAP}&\simeq& 1+2J_{0}^{2}(k(e)|\vec{r}_{i}-\vec{r}_{j}|) \\ &-&\frac{2}{A(\Omega)}\int J_{0}^{2}(k(e)|\vec{r}_{i}-\vec{r}|)d\vec{r} \nonumber \\ &-&\frac{2}{A(\Omega)}\int J_{0}^{2}(k(e)|\vec{r}'-\vec{r}_{j}|)d\vec{r}'\nonumber \\ &+&\frac{2}{A(\Omega)^{2}}\int \int J_{0}^{2}(k(e)|\vec{r}-\vec{r}'|)d\vec{r}d\vec{r}', \nonumber 
\end{eqnarray}
with $k(e)$ the wavenumber corresponding to the energy at the center of the spectral window.

Eq. (\ref{eq:Yun}), obtained using the Principle of Maximum Entropy and semiclassical considerations, is in fact the same result obtained using the supersymmetry technique in disordered systems \cite{mirl1}. The first line represents the usual Gaussian result, and the non-local terms, involving integrations over the whole space, are the normalization corrections up to second order in the inverse number of energy levels inside the spectral window.

We note that the theory of wavefunction's statistics in disordered systems, known as the diffusive sigma model, has a very different mathematical structure and physical motivation than our original idea of describing irregular eigenfunctions in single quantum systems. However, finding a way to use Berry's conjecture and/or the Principle of Maximum Entropy to derive a result coming from  the theory of disordered systems points towards a deep connection beyond the apparent differences between those fields. Research towards understanding the guiding principle behind this connection has been done for the universal fluctuations in the spectrum of statistical, chaotic and disordered systems \cite{sto1}, but not at the level of eigenfunctions. The present contribution goes into this direction, guided by Chladni's broad and elegant way of thinking.  
\subsection{Acknowledgments}

We gratefully acknowledge conversations with Georg Foltin, Sven Gnutzmann, Uzy Smilansky and Mark Dennis. We are indebted to Uzy Smilansky and Hans-J\"urgen St\"ockmann for organizing this excellent Chladni meeting. This work was started during a postdoctoral time of one of us (JDU) at the Weizmann Institute of Science supported by grant GIF 808/2003.

% If you prefer to produce your list of references by BIBTEX, please
% remove the thebibliography environment and include
% the following two commands instead:

% \bibliographystyle{epj}
% \bibliography{your bib files}

\begin{thebibliography}{}
\bibitem{berr1} M. V. Berry, J. Phys. A: Math. Gen. \textbf{10}, (1977) 2083.
\bibitem{sto1} The number of works testing the universal two-point correlation function is extensive. See for experimental results H. -J. St\"ockmann \textit{Quantum Chaos: An Introduction} (Cambridge University Press, Cambridge, England, 1999). For numerical results see for example A. B\"acker and R. Schubert, J. Phys. A: Math. Gen.  \textbf{35}, (2002) 527.

\bibitem{fredh}S. Fishman in \textit{Supersymmetry and Trace Formulae}, (Kluwer, New York, 1999) and references therein.

\bibitem{sied1}M. Sriednicki, Phys. Rev. E, \textbf{54}, (1996) 954, and \cite{mirl1} 

\bibitem{mirl1}I. V. Gornyi and A. D. Mirlin, Phys. Rev. E \textbf{65} (2002) 025202.

\bibitem{berr2} M. V. Berry, J. Phys. A: Math. Gen. \textbf{35}, (2002) 3025.

\bibitem{mirl2} A. D. Mirlin, Phys. Rep. \textbf{326}, (2000) 259.

\bibitem{KJD1}J. D. Urbina and K. Richter, Phys. Rev. Lett. \textbf{97}, (2006) 214101.

\bibitem{sied2} S. Hortikar and M. Sriednicki, Phys. Rev. Lett, \textbf{80}, (1998) 1646.

\bibitem{KJD2}J. D. Urbina and K. Richter, Phys. Rev. E. \textbf{70}, (2004) 015201(R).

\bibitem{berr3} M. V. Berry and I. Ishio, J. Phys. A: Math. Gen. \textbf{35} (2002) L447, W. E. Bies, N. Lepore and E. J. Heller, J. Phys. A: Math. Gen. \textbf{36} (2003) 1605.

\bibitem{rob} M. Robnik, J. Phys. A:Math. Gen. {\textbf 17} (1984) 1049.

\bibitem{KJD0}J. D. Urbina and K. Richter, J. Phys. A: Math. Gen. \textbf{36} (2003) L495.

\bibitem{brack}M. Brack and R. K. Badhuri, \textit{Semiclassical Physics} (Addison-Wesley, New York, 1997).

\bibitem{hell} E. J. Heller, in this conference and, E. J. Heller and B. R. Landry quant-ph/0704.188 (2007).

\bibitem{gold} S. Goldstein {\it et al}, J. Stat. Phys. (2004) and quant-ph/0309021.

\bibitem{PME} See, for example, W. Grandy \textit{Foundations of Statistical Mechanics} (D. Reidel Publishing Company, Holland, 1993).

\bibitem{JD3} J. D. Urbina, in preparation.  

% Format for book
% etc
\end{thebibliography}

\end{document}